\documentstyle[12pt]{article}

\author{B. Shklyar\\
Department of Mathematics and \\
Computer Science,\\
Bar-Ilan University,\\
Ramat-Gan 52900, Israel\\
shklyar@bimacs.cs.biu.ac.il}
\title{{\bf ON THE OBSERVABILITY FOR DISTRIBUTED SYSTEMS
BY MEANS OF LINEAR OPERATIONS}
}

\font\smb  = cmbx9

\newtheorem{theorem}{Theorem}
\newtheorem{lemma}{Lemma}

\newtheorem{definition}{Definition}

\begin{document}

\maketitle
\begin{abstract}
{\smb An observability problem for linear autonomous distributed systems in
the class of linear operations is considered. A criterion of observability
with respect to terminal state has been proved. A connection with
observability with respect to initial state is discussed.}
\end{abstract}

\section{Introduction}

One of main purposes of control systems is the determining the system state
functions. In order to construct feedback control for an optimal control
problem or for a stabilizability problem, the complete knowledge of the
state functions is required. But the system state functions may not be
directly measured, and often is it possible to obtain only some other
observations. Therefore the system state function should be determined from
the measured observed data. This problem is very important from theoretical
and practical points of view for theory of control systems.

The system state function can be uniquely determined from the initial state,
therefore determination of the initial state from the known observed
measured data assures determination of the state function. If the initial
state can be uniquely determined from the measured output data, the system
is said to be observable. Such observability for various classes of
distributed parameter systems is investigated in many publications (see, for
instance, \cite{1}-\cite{10}) for systems with aftereffect, for partial
parabolic and hyperbolic systems and generally for abstract systems in
Hilbert space with self-adjoint operator $A$. The observability criterion
for abstract systems in Hilbert space with a self-adjoint operator is
obtained in \cite{9}, sufficient observability conditions for abstract
systems in an arbitrary Banach space with a non-self-adjoint operator are
obtained in \cite{10}.

The present paper is devoted to a different observability conception. The
system is said to be observable, if the terminal state can be uniquely
determined from the measured output data. The approximated method of the
terminal state determination by means of a sequence of linear operation is
proposed and a criterion of existence of such sequence is obtained. The case
when it is possible to obtain initial state from terminal state, is
discussed.

\section{Problem statement}

Let $X$ be a Banach space, $Y=R^r,r\ge 1.$ Consider the equation
\begin{equation}
\label{e1}\dot x(t)=Ax(t),
\end{equation}
\begin{equation}
\label{e2}x(0)=x_0,
\end{equation}
\begin{equation}
\label{e3}y(t)=Cx(t),0\le t\le t_1,
\end{equation}
where $x(t)\in X$ is the current state, $x_0\in X$ is the initial state; $A$
is a linear operator whose domain $D(A)$ is dense in $X;A$ generates a
strongly continuous semi-group $S(t)$ of operators in the class $C_0$ \cite
{11}-\cite{12}; $y(t)\in Y;$ $C:X\rightarrow Y$ are linear operator with
domain $D(C)$ such that

(a) $x(t)\in D(C)\ {\rm for\ a.e.\ }t>0;$

(b) $y(.) \in L_2([0,t_1],Y), \forall t_1>0;$

(c) operator $Q:X\rightarrow L_2([0,t_1],Y), Qx=y(t), t \in [0,t_1]$ is
bounded for each $t_1>0.$

We consider only weak solutions \cite{12} of the above equation.

We assume $A$ to have the properties:

(i) the domain $D(A^{*})$ is dense in $X^{*}$;

(ii) the operator $A$ has a purely point spectrum $\sigma $ which is either
finite or has no finite limit points and each $\lambda \in \sigma $ has a
finite multiplicity;

(iii) there is a time $T\ge 0$ such that for each $x_{0}\in X$ and $t>T$ the
function $x(t)=S(t)x_{0}$ is expanded in a series of eigenvectors and
associated vectors of $A$, converging uniformly with respect to $t$ on an
arbitrary interval $[T_{1},T_{2}],T_{2}>T_{1}>T$ for a certain grouping of
terms.

Denote by $\sigma $ the spectrum of operator $A$. If $x\in X$ and $f\in
X^{*} $, we will write $(x,f)$ instead $f(x)$. The superscript $T$ denotes
transposition.

We denote by ${\bf C}$ the complex plane.

Together with equation (\ref{e1})-(\ref{e2}) we consider the equation
\begin{equation}
\label{e4}\dot x(t)=A^{*}x^{*}(t)+C^{*}u(t),0\le t\le t_1,
\end{equation}
\begin{equation}
\label{e5}x^{*}(0)=x^{*},
\end{equation}
where $x^{*}\in X^{*},u\in Y^{*}.$

The function $x^{*}:R^1\rightarrow X^{*}$, where $x^{*}(t)$ is linear
bounded functional, generating by means of expression
\begin{equation}
\label{e6}(x,x^{*}(t))=(S(t)x,x^{*})+\int_0^ty^T(t-\tau )u(\tau )d\tau
\end{equation}
is said to be solution of equation (\ref{e4}) with initial condition (\ref
{e5}).

\section{Preliminary results}

In order to obtain the principal result of this paper, we shall prove a
number of additional results.

The sequence $x_{i}(t), i=1,2,\ldots $ of functions from $L_{2}[0,t_{1}]$ is
called minimal, iff there is a sequence $y_{j}(t), i=1,2,\ldots ,$ of
functions from $L_{2}[0,t_{1}]$, such that
$$
\int^{t_{1}}_{0}(x^{T}_{i}(t)y_{j}(t))\,dt = \delta _{ij}, i,j=1,2,\ldots,
$$
where $\delta _{ij}$ is the Kronecker's symbol. The sequence $y_{j}(t),
j=0,1,\ldots $ is called biorthogonal to the sequence $x_{j}(t),
j=1,2,\ldots $.

Let numbers $\lambda _j\in \sigma ,j=1,2,\ldots $ be enumerated in the order
of non decreasing absolute values, let $\alpha _j$ be the multiplicity of
$\lambda _j\in \sigma $, and let
$\varphi _{ij},i=1,2,\ldots ,j=1,2,\ldots ,\beta _i,\beta _i\le \alpha _i$,
$A\varphi _{i\beta _i}=\lambda _i\varphi_{i\beta _i}$ be the root vectors
(eigenvectors and associated vectors) of the operator
$A;\psi _{kl},k=1,2,\ldots ,l=1,2,\ldots ,\beta _k$, be the
root vectors of the adjoint operator $A^{*}$, such that
\begin{equation}
\label{e7}
(\varphi _{p\beta _p-l+1,}\psi _{jk})=\delta _{pj}\delta_{lk},
p,j=1,2,\ldots ,l=1,\ldots ,\beta _p,k=1,\ldots ,\beta _j.
\end{equation}

We use the following notations:
$$
c_{ij}=C\varphi _{ij}, i=1,2,\ldots ,j=1,2,\ldots \beta _{i};
I_{i}=\{j:\lambda _{j}=\lambda _{i}\},
$$
$\gamma _{i}$ is the number of elements in $I_{i},i=1,2,\ldots $ ;
$$
f_{jk}(t)=t^{k-1}\exp (\lambda _{j}t), j=1,2,\ldots , k=1,2,\ldots ,\alpha
_{j}, t\in [0,t_{1}],
$$
$$
g_{jk}(t)= \exp (\lambda _{j}t)\sum^{\beta _{j}-k}_{l=0} c_{jk+l}{\frac{%
t^{l} }{j!}}, j=1,2,\ldots , k=1,\ldots ,\beta _{j}, t\in [0,t_{1}].
$$

\begin{lemma}
If the sequence \label{L1}
\begin{equation}
\label{e8}f_{jk}(-t),k=1,2,\ldots ,\alpha _j,t\in [0,t_1]
\end{equation}
is minimal;
\begin{equation}
\label{e9}Ker(Ax-\lambda x)\bigcap KerC=0\ {\rm for\ each\ }\lambda \in
\sigma ,
\end{equation}
then the sequence
\begin{equation}
\label{e10}g_{jk}(-t),j=1,2,\ldots ,k=1,\ldots ,\beta _j,0\le t\le t_1,
\end{equation}
is minimal.
\end{lemma}

{\it Proof}. Let $u_{s\nu }(t),\nu =0,1,\ldots ,\alpha _s-1,s=1,2,\ldots $
be the sequence, biorthogonal to the sequence (\ref{e8}). Consider the
sequence
\begin{equation}
\label{e11}v_{sp}(t)=\sum_{\nu =0}^{\alpha _s-1}\xi _{sp\nu }u_{s\nu
}(t),s=1,2,\ldots ,p=1,2,\ldots ,\alpha _s,
\end{equation}
where $\xi _{sp\nu }\in Y$. We'll chose the $\xi _{sp\nu }$ so that sequence
(\ref{e11}) is biorthogonal to sequence (\ref{e10}). We have

\begin{equation}
\label{e12}
\int_0^{t_1}g_{jk}^T(-t)v_{sp}(t)dt=
\cases{0,&if $j\in I_\mu ,s\in I_\eta ,\mu \neq \eta$,\cr
\sum_{\nu =0}^{\beta _j-k}c_{jk+\nu }^T\xi _{sp\nu },&if $j\in I_\mu ,s\in
I_\eta ,\mu =\eta$.\cr}
\end{equation}
Sequence (\ref{e11}) is biorthogonal to sequence $g_{jk}(-t)$, iff
\begin{equation}
\label{e13}
\sum_{\nu =0}^{\beta _j-k}c_{jk+\nu }^T\xi _{sp\nu }=\delta_{js}\delta _{kp},
\end{equation}
$$
s,j\in I_\mu ,k=1,\ldots \beta _j,p=1,\ldots ,\beta _s.
$$
For $k=\beta _j$ we have
\begin{equation}
\label{e14}c_{j\beta _j}^T\xi _{sp0}=\delta _{js}\delta _{\beta _jp},s,j\in
I_\mu ,p=1,\ldots ,\beta _s.
\end{equation}
It follows from (\ref{e9}) that vectors $c_{j\beta _j},j\in I_\mu $, are
linearly independent and $\gamma _\mu \le ${\it r}. Hence the linear
algebraic system (\ref{e14}) of $\gamma _\mu $ equations for the $r$ unknown
components of $\xi _{sp0}$ has a solution for $j,s\in I_\mu $ and $%
p=1,\ldots ,\beta _s$. For $k=\beta _j-1$ we have
\begin{equation}
\label{e15}
c_{j\beta _j-1}^T\xi _{sp0}-c_{j\beta _j}^T\xi _{sp1}=
\delta_{js}\delta _{\beta _j-1,p},s,j\in I_\mu ,p=1,\ldots ,\beta _s.
\end{equation}
Using (\ref{e14}) for the vectors
$\xi _{sp0},s\in I_\mu ,p=1,\ldots ,\beta_s$
in (\ref{e15}), we obtain a linear algebraic system of $\gamma _\mu $
equations for the $r$ unknown components of vector $\xi _{sp1},s\in I_\mu,
p=1,\ldots ,\beta _s,\mu =1,2,\ldots $ . By (\ref{e9}) this system has a
solution. Using (\ref{e13}) for $k=\beta _j-l$, where
$l=1,2,\ldots ,\beta_j $, we conclude that, for known
$s\in I_\mu ,\nu =0,1,\ldots ,l-1,$ the system (\ref{e15}) is a linear
system of $\gamma _\mu $ equations for the $r$
unknown components of vector $\xi _{spl}$, having a solution by (\ref{e9}).
This proves the lemma.

\begin{definition}
\label{D0} A vector $x^{*}\in X^{*}$ is said to be completely controllable
on $[0,t_1]$, if there is a control $u(t),u(.)\in L([0,t_1],Y^{*})$, such
that
\begin{equation}
\label{e17}(S(t_1)x,x^{*})+\int_0^{t_1}y^T(t_1-\tau )u(\tau )d\tau =0
\end{equation}
for each $x\in X$.
\end{definition}

\begin{lemma}
\label{L2} For $x^{*}\in X^{*}$ to be completely controllable on $[0,t_1]$,
it is necessary that the moment problem (finite or infinite)
\begin{equation}
\label{e16}
(\varphi _{jk},x^{*})=\int_0^{t_1}g_{jk}^T(-t)u(t)dt,j=1,2,\ldots,
k=1,2,\ldots ,\beta _j
\end{equation}
has a solution $u^0(.)\in L_2^r[0,t_1]$. If the moment problem (\ref{e16})
has a solution, then $x^{*}$ is completely controllable on $[0,t_1+T]$.
\end{lemma}

{\it Proof}. {\it Necessity}. If $x^*$ is completely controllable on
$[0,t_{1}]$, then in virtue of (\ref{e6}) there exists $u(t)\in
L^{r}_{2}[0,t_{1}]$ such that (\ref{e17}) holds. Substituting $\varphi_{jk},
j=1,2,\ldots , k=1,2,\ldots ,\beta _{j}$ to both parts of (\ref{e17})
instead $x$, we obtain (\ref{e16}).

{\it Sufficiency}. Let $x_{jk}(t)=(\varphi _{jk},x(t)),j=1,2,\ldots,
k=1,2,\ldots ,\beta _j$, where $x(t)$ is the weak solution of equation
(\ref{e4}) with initial condition (\ref{e5}). It easy to show that the
sequence $x_{jk}(t)$ is the solution of the infinite system
$$
\dot x(t)=\lambda _jx_{jk}(t)+x_{jk+1}(t)+c_{jk}u(t),
k=1,2,\ldots ,\beta_{j-1};
$$
\begin{equation}
\label{e18}\dot x(t)=\lambda _j{x_{j\beta _j}}(t)+c_{j\beta _j}u(t),
\end{equation}
$$
j=1,2,\ldots
$$
with initial conditions
$$
x_{jk}(0)=(\varphi _{jk},x^{*}),j=1,2,\ldots ,k=1,2,\ldots ,\beta _j.
$$
Let $u^0(t),t\in [0,t_1],u^0(.)\in L_2^r[0,t_1]$ be a solution of moment
problem (\ref{e16}). If $u(t)=u^0(t)$ a.e. on $[0,t_1]$ than by (\ref{e16})
and (\ref{e18}) we have
$$
x_{jk}(t_1)=0,j=1,2,\ldots ,k=1,2,\ldots ,\beta _j.
$$
Putting $u(t)\equiv 0,t\ge t_1$, we conclude from (\ref{e18}) that
$$
x_{jk}(t)\equiv 0,j=1,2,\ldots ,k=1,2,\ldots ,\beta _j,\forall t\ge t_1,
$$
and $x^{*}(t)$ is the solution of the equation
$$
\dot x(t)=A^{*}x^{*}(t)
$$
for each $t\ge t_1$. By property (iii) of the operator $A$, we obtain
$$
x^{*}(t)\equiv 0,\forall t>t_1+T.
$$
This proves the lemma.

\section{Main results}

In this section we will obtain the observability criterion by means of
linear operation and show a way to restore the $x(t_{1})$ by output
(\ref{e3}).

Denote by $L(Y,X)$ the space of linear bounded operators acting from $Y$
to $X$.

\begin{definition}
\label{D1} The equation (\ref{e1})-(\ref{e3}) is called approximately
observable on $[0,t_1]$ in the class of linear operations, if for any
$\epsilon >0$ and each solution $x(t)$ of equation (\ref{e1})-(\ref{e2})
there exists an operator-valued function
$$
U_{\epsilon x}(t),0\le t\le t_1,U_{\epsilon x}\in L_2([0,t_1],L(Y,X))
$$
such that
\begin{equation}
\label{e19}\Vert x(t_1)-\int_0^{t_1}U_{\epsilon x}(\tau )y(t_1-\tau )d\tau
\Vert <\epsilon .
\end{equation}
\end{definition}

\begin{theorem}
\label{T1} Let sequence (\ref{e8}) be minimal. Equation (\ref{e1})-(\ref{e2}%
) is approximately observable on $[0,t_1+T]$ in class of linear operations,
iff (\ref{e9}) holds.
\end{theorem}

{\it Proof}. {\it Necessity}. Assume that (\ref{e9}) does not hold. Then
there exists $a \lambda \in \sigma $ and vector $\varphi _{\lambda } \in X,
\varphi _{\lambda } \neq 0$ such that
$$
C\varphi _{\lambda }=0, A\varphi _{\lambda }-\lambda \varphi _{\lambda }=0.
$$
Hence $x_{\lambda }(t)=\varphi _{\lambda }\exp (\lambda t)$ is the solution
of equation (\ref{e1}) such that $Cx_{\lambda }(t)\equiv 0, \forall t\ge 0$
and so
$$
\|x_{\lambda }(t)-\int^t_0 U_{\epsilon x}(\tau) Cx_{\lambda }(t-\tau)d\tau
\|= \| \varphi_{\lambda } \| \exp (\lambda t)
$$
for any operator-valued function
$$
U_{\epsilon x}(\tau), 0\le \tau\le t,U_{\epsilon x}(.)\in
L_{2}([0,t],L(Y,X))
$$
This proves the necessity.

{\it Sufficiency}. By Lemma \ref{L1} and (\ref{e7}) there exists functions
$$
u_{jk}(t),j=1,2,\ldots ,k=1,2,\ldots ,\beta _j
$$
such that
\begin{equation}
\label{e20}
(\varphi _{p\beta _p-l+1},\psi _{jk})=
\int_0^{t_1}(g_{p\beta_p-l+1}^T(-t)u_{jk}(t))dt,
\end{equation}
$$
p=1,2,\ldots ,l=1,2,\ldots ,\beta _p,
$$
hence by Lemma \ref{L2} the $\psi _{jk},j=1,2,\ldots ,k=1,2,\ldots ,\beta _j$
are completely controllable on $[0,t_1+T]$. According to (\ref{e20}) and
Definition \ref{D0} we obtain
\begin{equation}
\label{e26}(x(t_1+T),\psi _{jk})+\int_0^{t_1}(Cx(t_1+T-\tau ))^Tu_{jk}(\tau
)d\tau =0
\end{equation}
for each solution $x(t)$ of equation (\ref{e1}). By property (iii) of
operator $A$ we have
\begin{equation}
\label{e27}
x(t)=\sum_{j=1}^\infty \sum_{k=1}^{\beta _j}(x(t),\psi_{jk})\varphi _{jk},
\forall t>T.
\end{equation}
Using (\ref{e26}) and (\ref{e27}), we conclude that

$x(t_1+T)=$
\begin{equation}
\label{e28}
\sum_{j=1}^\infty \sum_{k=1}^{\beta _j}\varphi_{jk}
\int_0^{t_1}u_{jk}^T(\tau )y(t_1+T-\tau )d\tau )=\lim _{p\rightarrow
\infty }\int_0^{t_1}U_p(\tau )y(t_1+T-\tau )d\tau ,
\end{equation}
where $U_p(\tau ):R^r\rightarrow X$ is the linear bounded operator, defined
by formula
\begin{equation}
\label{e29}
U_p(\tau )y=
-\sum_{j=1}^p\sum_{k=1}^{\beta _j}(\varphi_{jk}u_{jk}^T(\tau ))y.
\end{equation}
The theorem has been proved.

We can conclude from (\ref{e28}) that we use information about behavior of
output $y(t)$ only on $[T,t_{1}+T]$ in order to determine $x(t_{1}+T)$. If
$t_{1}>T$, then by means of more precise investigation of properties of
attainable set for equation (\ref{e4}) and by using information of behavior
of $y(t)$ on all $[0,t_{1}]$ it is possible to determine terminal state
$x(t_{1})$.

Let $K(t)$ is the attainable set of equation (4), that is
$$
K(t)=\{x^{*}:x^{*}=x^{*}(t),u(.)\in L_2^r[0,t_1]\},
$$
where $x^{*}(t)$ is the solution of equation (\ref{e4}) with initial
condition $x^{*}(0)=0$, which corresponds to control $u(t)$. It is easy to
show, that $x^{*}\in K(t)$ iff there exists a control $u(t),u(.)\in
L_2^r[0,t_1]$, such that
\begin{equation}
\label{e291}
(x,x^{*})=\int_0^{t_1}y^T(t_1-\tau )u(\tau )d\tau ,\forall x\in X.
\end{equation}

\begin{lemma}
\label{L3} If $t_1\le t_2$, then $K(t_1)\subseteq K(t_2)$.
\end{lemma}

{\it Proof}. Let $t_{1}<t_{2}$ and $x^*\in K(t_{1})$, i.e. there is a
control $u_{1}(.)\in L^{r}_{2}[0,t_{1}]$, such that (\ref{e291}) holds.

Let
$$
u_2(\tau )=
\cases{0,&if $0\leq \tau \leq t_{2-}t_{1},$\cr
u_1(t_{1}-t_2+\tau ),&if $t_{2-}t_1\leq \tau \leq t_2$.\cr}
$$
It is easy to show, that
$$
(x,x^{*})=\int_0^{t_2}y^T(t_2-\tau )u_2(\tau )d\tau ,
$$
and $u_2(.)\in L_2^r[0,t_2].$ Hence, $K(t_1)\subseteq K(t_2)\ {\rm for\ all\
}t_1<t_2.$

\begin{theorem}
\label{T2} Let sequence (\ref{e8}) be minimal, $t_1>T$ and $K(t+T)\equiv
K(t),\forall t\ge T$.

Equation (\ref{e1})-(\ref{e2}) is approximately observable on $[0,t_1]$ in
class of linear operations, iff (\ref{e9}) holds.
\end{theorem}

{\it Proof}. One should to prove only sufficiency. Let $A_j$ be
$(\beta_j\times \beta _j)$ Jordan matrix;

$$
A_j=\left\{
\begin{array}{cccc}
\lambda _j & 0 & ... & 0 \\
1 & \lambda _j & ... & 0 \\
... & ... & ... & ... \\
0 & 0 & ... & \lambda _j
\end{array}
\right\}
$$

We denote by $X_j(t)$ the matrix
$\{x_{j1}(t),x_{j2}(t),\ldots ,x_{j\beta_j}(t)\},j=1,2,\ldots $,
where $x_{jk}(t)$ is the solution of equation (\ref{e4}) with initial
condition $x_{jk}(0)=\psi _{jk}$, corresponding to the
control $u_{jk}(t),0\le t\le t_1$ and let $U_j(t)$ be the matrix
$\{u_{j1}(t),u_{j2}(t),\ldots ,u_{j\beta _j}(t)\},j=1,2,\ldots $.
Denote by $(x,X_j(t))$ the row
$\{(x,x_{j1}(t)),\cdots ,(x,x_{j\beta _j(t)})\}$.
It follows from (\ref{e6}) that
$$
(x,X_j(t))=(x,X_j(0)\exp (A_j(t))+\int_0^{\min (t,t_1)}y^T(t-\tau )U_j(\tau
)d\tau ,\forall x\in X
$$
and from (\ref{e26}) we have
\begin{equation}
\label{e21}(x,X_j(0)\exp (A_jt))+\int_0^{t_1}y^T(t+T-\tau )U_j(\tau )\exp
(-A_jT)d\tau =0,
\end{equation}
$\forall t>t_1,\forall x\in X.$

Since
$$
\int_0^ty^T(t+T-\tau )U_j(\tau )\exp (-A_jT)d\tau \in
(x,K(t+T)XK(t+T)X\ldots XK(t+T))
$$
and $K(t+T)\equiv K(t),\forall t>T$, there exist controls $%
v_{jk}(t),j=1,2,\ldots ,k=1,2,\ldots ,\beta _j,$ such that for all $t>T$
\begin{equation}
\label{e30}\int_0^ty^T(t+T-\tau )U_j(\tau )\exp (-A_jT)d\tau=
\int_0^{t_1}y^T(t-\tau )V_j(\tau )d\tau ,\forall x\in X,
\end{equation}
where $V_j(t)=\{v_{j1}(t),v_{j2}(t),\ldots ,v_{j\beta _j}(t),j=1,2,\ldots .$
Substituting (\ref{e30}) into (\ref{e21}) we obtain that the
$\psi_{jk},j=1,2,\ldots ,k=1,2,\ldots ,\beta _j$ are completely
controllable on $[0,t_1],t_1>${\it T}.
Using value $t_1$ instead value $t_1+T$ in formulas
(\ref{e26})-(\ref{e29}), we prove the theorem.

{\it Remark}. If the sequence $\psi _{jk}, j=1,2,\ldots , k=1,2,\ldots \beta
_{j}$ is complete, that is
$$
(x,\psi _{jk})=0, j=1,2,\ldots , k=1,2,\ldots \beta _{j}
$$
iff $x=0$, then the Theorem 3 from \cite{10} immediately follows from
theorems \ref{T1}-\ref{T2}. Moreover, it is possible to determine the
terminal state $x(t_{1})$ of equation (\ref{e1}) by means of formulas (\ref
{e11}), (\ref{e20}), (\ref{e28}), (\ref{e29}) and then to determine the
initial state $x_{0}$ of equation (\ref{e1}), using the completeness of
vectors $\psi _{jk}, j=1,2,\ldots , k=1,2,\ldots \beta _{j}$ and properties
of operator $A, C$ for a given specific equation (\ref{e1})-(\ref{e3}).

\begin{definition}
\label{D2} The equation (\ref{e1})-(\ref{e3}) is said to be observable on $%
[0,t_1]$ if initial state $x_0$ is uniquely determined by output (\ref{e3}),
$0\le t\le t_1$.
\end{definition}

\begin{definition}
\label{D3} The equation (\ref{e1}) is said to be pasted on $[0,t_2]$, if
there is $x_0\in X$, such that $S(t)x_0\equiv 0,t_2\le t<\infty $.
\end{definition}

\begin{definition}
\label{D4} The equation (\ref{e1}) is said to be non-pasted, if there no $%
t_2,0<t_2<\infty $, such that equation (\ref{e1}) is pasted on $[0,t_2]$.
\end{definition}

It follows from Theorems \ref{T1}-\ref{T2} that next theorem is true.

\begin{theorem}
\label{T3} If (\ref{e9}) holds, sequence (\ref{e8}) is minimal and equation (%
\ref{e1}) is non-pasted, than equation (\ref{e1})-(\ref{e3}) is observable
on $[0,t_1+T]$. If $t_1>T$ and $K(t+T)\equiv K(t),t\ge T$, then equation (%
\ref{e1})-(\ref{e3}) is observable on $[0,t_1]$.
\end{theorem}

By Theorem \ref{T3}, the terminal state $x(t_{1})$ can be uniquely
determined by means of linear operations from measured output (\ref{e3}), if
(\ref{e9}) holds and sequence (\ref{e8}) is minimal. If equation (\ref{e1})
is non-pasted, then the initial state $x_{0}$ can be uniquely determined
from measurements of $x(t_{1})$. The problem of initial state determination
is not necessarily well-posed \cite{9}. For self-adjoint operator $A$ it is
possible to apply an approximate method which reduces this ill-posed problem
to a well-posed one \cite{9}.

\section{Examples}

Analyzing specific spaces $X$ and operators $A$ and $C$ it is possible to
obtain results on the observability problems (both previously known and new
ones) for dynamical systems described, for example, by partial differential
parabolic and hyperbolic equations, systems with aftereffect,
integro-differential equations and other equations, satisfying the
conditions (i), (ii), (iii), and for which the sequence (\ref{e8}) is
minimal.

A number of conditions for sequence (\ref{e8}) to be minimal are proved in
\cite{13}-\cite{16}.

Let $X$ be a Hilbert space. Consider equation (\ref{e1})-(\ref{e3}) with
self-adjoint operator $A$. It is well-known that in this case the spectrum
$\sigma $ of $A$ is the sequence $\{\lambda _j,j=1,2,\ldots \}$ of real
negative numbers; $\lim _{j\to \infty }{\lambda _j}=-\infty $; operator $A$
has properties (i)-(iii) with $T=0$. Let $\varphi _{jk},j=1,2,\ldots,
k=1,2,\ldots ,\gamma _j$ be the eigenfunctions of $A$ corresponding to
eigenvalues $\lambda _j,j=1,2,\ldots $. It is well-known, that the sequence
$\varphi _{jk},j=1,2,\ldots ,k=1,2,\ldots ,\gamma _j$ is complete. It is easy
to prove, that in this case condition (\ref{e9}) is equivalent to linear
independence of vectors
$$
C\varphi _{jk},k=1,2,\ldots ,\gamma _j
$$
for all $j,j=1,2,\ldots ,$ and it follows from results of \cite{16} that
sequence of real exponents
$$
\exp (-\lambda _jt),0\le t\le t_1,\forall t_1>0
$$
is minimal
\footnote {In \cite{16} the sequence of positive numbers
$\lambda_j, j=1,2,\ldots$ is considered, but the corresponding results
of \cite{16} are easy applied to the sequence $\sigma$ of negative
numbers $\lambda_j, j=1,2,\ldots$} provided that
$$
\sum_{j=1}^\infty \frac 1{|\lambda _j|}<\infty .
$$
Hence equation (1)-(3) is approximately observable in the class of linear
operations for any $t_1>0$, iff the vectors
$$
C\varphi _{jk},k=1,2,\ldots ,\gamma _j
$$
are linearly independent for all $j,j=1,2,\ldots .$

Since the sequence of eigenfunctions of $A$ is complete, the equation (\ref
{e1})-(\ref{e3}) is observable on $[0,t_1]$, if last condition holds. Thus,
all results from \cite{2},\cite{8},\cite{9}, concerning observability
criteria, are the immediate corollaries of Theorem \ref{T1}. Moreover it is
possible to restore $x(t,\xi)$ by means of linear operations for any $t>0$.

Consider for instance the one-dimensional heat equation
\begin{equation}
\label{e301}{\frac{\partial x}{\partial t}}(t,\xi )={\frac{\partial ^2x}{%
\partial \xi ^2}}(t,\xi );
\end{equation}
\begin{equation}
\label{e302}x(0,\xi )=\varphi (\xi ),0\le \xi \le \pi ,
\end{equation}
\begin{equation}
\label{e303}x(t,0)=0,x(t,\pi )=0,0\le t<t_1,
\end{equation}
\begin{equation}
\label{e304}y(t)=\int_0^\pi b(\xi )x(t,\xi )d\xi ,
\end{equation}
where $x,\varphi ,b,y\in R^1,\varphi (.),b(.)\in L_2[0,\pi ].$

The equation (\ref{e301}) is the particular case of the problem (\ref{e1})-(%
\ref{e3}), where
$$
X=L_2[0,\pi ], (Ax)(\xi )={\frac{d^2x }{d\xi^2}}(\xi )
$$
with domain
$$
D(A)=\{ x \in C^2[0,\pi]: x(0)=x(\pi)=0 \};
$$
$$
Cx=\int^{\pi }_0 b(\xi)x(\xi) d\xi.
$$
Problem (\ref{e301})-(\ref{e303}) is uniformly well-posed, hence operator $A$
generates a $C_{0}$-semigroup;
$\sigma =\{-j^{2},j=1,2,\ldots \};
\alpha_{j}=\beta _{j}=1, \varphi _{j}(\xi )=\sin (jx)$
are the eigenfunctions of operator $A$, corresponding to eigenvalues
$\lambda _{j}=-j^{2},j=1,2,\ldots $; $A$ is self-adjoint operator;
for each $\varphi (.)\in X$
the corresponding solution $x(t,\xi )$ of equation (\ref{e301})-(\ref{e303})
is expanded into a series
$$
x(t,\xi )= \sum^{\infty }_{j=1}(\int^{\pi }_{0}\varphi (\xi )\sin (j\xi
)d\xi ) \exp (-j^{2}t),
$$
convergent uniformly for any segment $[0,h]$, that is in the given case we
have $T=0$; $\sum^{\infty}_{j=1}\frac{1}{j^2} < \infty$, so it follows from
results of \cite{16} that sequence $\exp (j^{2}t), 0\le t\le t_{1}$ is
minimal for any $t_{1}>0.$

\begin{definition}
\label{D5} Equation (\ref{e301})-(\ref{e304}) is said to be approximately
observable on $[0,t_1]$ in the lass of linear operations, if for any $%
\epsilon >0$ and for each solution $x(t,\xi )$ of equation (\ref{e301})-(\ref
{e303}) there exists a function $u_{\epsilon x}(t,\xi ),0\le t\le
t_1,u_{\epsilon x}(t,.)\in L_2[0,\pi ]$, such that
$$
\int_0^{t_1}|x(t_1,\xi )-\int_0^\pi u_{\epsilon x}(\tau ,\xi )y(t_1-\tau
)d\tau |^2d\xi <\epsilon .
$$
\end{definition}

Condition (\ref{e9}) for equation (\ref{e301})-(\ref{e304}) is equivalent to
condition
\begin{equation}
\label{e305}\int_0^\pi b(\xi )\sin (j\xi )d\xi \neq 0
\end{equation}
for each $j=1,2,\ldots $. In accordance with Theorem \ref{T2} we conclude
that equation (\ref{e301})-(\ref{e304}) is approximately observable on $%
[0,t_1],t_1>0$ in the class of linear operations, iff (\ref{e305}) holds.

Characteristic matrix of equation (\ref{e301})-(\ref{e303}) are of the form
\begin{equation}
\label{e306}\left\{
\begin{array}{cc}
1 & 1 \\
\exp (-\pi \sqrt{\lambda }) & \exp (\pi \sqrt{\lambda }
\end{array}
\right\}
\end{equation}
Using (\ref{e306}), it is possible to write (\ref{e305}) in another
equivalent form
\begin{equation}
\label{e307}{\rm rank}\left\{
\begin{array}{ccc}
1 & \exp (-\pi \sqrt{\lambda }) & \int_0^\pi b(x)\exp (-\pi
\sqrt{\lambda }x)dx \\ 1 & \exp (\pi \sqrt{\lambda } & \int_0^\pi b(x)\exp
(\pi \sqrt{\lambda }x)dx
\end{array}
\right\} =2
\end{equation}
for all complex $\lambda $.

In difference from (\ref{e305}) we should not compute eigenvectors of
equation (\ref{e301})-(\ref{e303}) in order to verify the (\ref{e307}) (this
problem is trivial for equation (\ref{e301})-(\ref{e303}), but it can be
non-trivial for more complicated equations).

Now we will consider an important example of equation (\ref{e1})-(\ref{e3})
with non-self-adjoint operator $A$ and with $T>0.$

Consider the equation
\begin{equation}
\label{e31}
{\frac{\partial x}{\partial t}}(t,\xi )={\frac{\partial x}
{\partial \xi }}(t,\xi ),0\le t\le t_1,-h\le \xi \le 0,
\end{equation}
\begin{equation}
\label{e32}
\sum_{j=0}^mA_{0j}{\frac{\partial x}{\partial \xi }}
(t,-h_j)+A_{1j}x(t,-h_j)=0,
\end{equation}
\begin{equation}
\label{e33}x(0,\xi )\equiv \varphi (\xi ),-h\le \xi \le 0,
\end{equation}
\begin{equation}
\label{e34}y(t)=\sum_{j=0}^mB_jx(t,-h_j),0\le t\le t_1,
\end{equation}
where $x\in R^n,y\in R^r,A_{0j}$ and $A_{1j}$ are constant $nXn$-matrices,
$B_j$ are constant $rXn$-matrices, $0=h_0<h_1<\ldots <h_m=h$ are the real
numbers, $\varphi (\xi ),-h\le \xi \le 0$ is an absolutely continuous
$n$-vector-valued function. It is easy to show \cite{12}, that problem (\ref
{e31})-(\ref{e34}) is a particular case of problem (\ref{e1})-(\ref{e3}),
where the corresponding operator $A$ has the properties (i)-(iii) with $T=nh$.

\begin{definition}
\label{D6} Equation (\ref{e31})-(\ref{e34}) is said to be approximately
observable on $[0,t_1]$ in the class of linear operations, if for any $%
\epsilon >0$ and for each solution $x(t,\xi )$ of equation (\ref{e31})-(\ref
{e34}) there exists an $nXr$-matrix-function $U_{\epsilon x}(\tau ,\xi ),$

$U_{\epsilon x}(.)\in L_2^{nr}[0,t_1]XL_2^{nr}[-h,0]$, such that
$$
\Vert x(t_1,\xi )-\int_0^{t_1}U_{\epsilon x}(\tau ,\xi )y(t_1-\tau )d\tau
\Vert <\epsilon ,%
\footnote {In this case symbol
$\|.\|$ denotes a norm in $R^n.$}\ \forall \xi \in [-h,0].
$$
\end{definition}

\begin{theorem}
\label{T4} For the system (\ref{e31})-(\ref{e34}) to be approximately
observable on $[0,t_1]$ by means of linear operations, it is necessary, and
for $t_1>2nh$ sufficient, that relation holds for

\begin{equation}
\label{e35}{\rm ra}{\rm nk}\{\sum_{j=0}^mA_{0j}^T\lambda +A_{1j}^T)\exp
(-\lambda h_j),\sum_{j=0}^mB_j^T\exp (-\lambda h_j)\}=n
\end{equation}

holds for all complex $\lambda $.
\end{theorem}

{\it Proof}. For the system (\ref{e31})-(\ref{e34}) we have \cite{12}
\begin{equation}
\label{e36}(Ay)(\tau )=\dot y(\tau ),-h\le \tau \le 0,
\end{equation}
$D(A)=$
\begin{equation}
\label{e37}=\{x(.)\in C^n[-h,0],\dot x(.)\in
C^n[-h,0],\sum_{j=0}^mA_{0j}\dot x(-h_j)+A_{1j}x(-h_j)=0\},
\end{equation}
\begin{equation}
\label{e38}Cx=\sum_{j=0}^mB_jx(-h_j),0\le t\le t_1.
\end{equation}
It is easy to shown, that equalities
$$
Ax-\lambda x=0,Cx=0
$$
for the operators $A$ and $B$, defined by (\ref{e36})-(\ref{e38}), are
equivalent to the equalities:
$$
(\sum_{j=0}^mA_{0j}\lambda +A_{1j})\exp (-\lambda h_j))x(0)=0;
$$
\begin{equation}
\label{e39}(\sum_{j=0}^mB_j\exp (-\lambda h_j))x(0)=0;
\end{equation}
Hence, condition (\ref{e9}) is equivalent to condition (\ref{e35}), and it
follows from (\ref{e39}) that the spectrum of the system (\ref{e31})-(\ref
{e32}) is the set
$$
\{\lambda \in {\bf C}:\det \sum_{j=0}^m(A_{0j}\lambda +A_{1j})\exp (-\lambda
h_j)=0\}.
$$
It is proved in \cite{17}, that in this case lemma \ref{L1} is true for $%
t_1>nh$. The theorem follows from Theorem \ref{T1}.

The general solution $x(t,\xi )$ of the equation (\ref{e31}) is defined by
the formula: $x(t,\xi )=z(t+\xi )$, where $z(t)$ is an arbitrary absolutely
continuous $n$-vector-valued function. Hence, $x(t,\xi )$ is the solution of
equation (\ref{e31}) with boundary condition (\ref{e32}), iff $z(t)$ is the
solution of the neutral system \cite{18},\cite{19}
\begin{equation}
\label{e40}\sum_{j=0}^mA_{0j}\dot x(t-h_j)+A_{1j}x(t-h_j)=0,
\end{equation}
and observability problem (\ref{e31})-(\ref{e34}) is equivalent to the
observability problem for equation (\ref{e36}) with the output
\begin{equation}
\label{e41}y(t)=\sum_{j=0}^mB_jx(t-h_j),0\le t\le t_1.
\end{equation}
This problem was investigated in \cite{17} for the case when $%
B_0=B,B_j=0,j=1,\ldots ,m$, and Theorem \ref{T4} was obtained for this case
by a different method
\footnote {A matrix-valued function
$U_{\epsilon x}(\tau,\xi)$ can be written in the form:
$U_{\epsilon x}(\tau,\xi)=V_{\epsilon x}(\tau+\xi)$, where elements
$v_{ij}(t), i=1,...,n, j=1,...,r$ of matrix $V_{\epsilon x}(t)$
belong to $L_2[0,t_1]$ and can be calculated by means of explicit formulas
\cite{17}. }.

\begin{theorem}
\label{T5} For the equation (\ref{e31})-(\ref{e32}) to be non-pasted, it is
necessary and sufficient, that there exists $\lambda \in {\bf C}$, such that
\begin{equation}
\label{e42}{\rm rank}\{A_{0m}\lambda +A_{1m}\}=n.
\end{equation}
\end{theorem}

{\it Proof}. {\it Sufficiency}. We have by (\ref{e36}) that $S(t)x_0\equiv
0,t\ge t_2$, iff
\begin{equation}
\label{e43}\sum_{j=k+1}^mA_{0j}\dot x(t-h_j)+A_{1j}x(t-h_j)\equiv 0,
\end{equation}
$$
t_2+h_k<t\le t_2+h_{k+1},k=0,1,\ldots ,m-1;x(t_2)=0.
$$
If $k=m-1,$ we have by (\ref{e43}):
\begin{equation}
\label{e44}A_{0m}\dot x(t-h_m)+A_{1m}x(t-h_m)\equiv 0,t_2+h_{m-1}<t\le
t_2+h_m.
\end{equation}
If (\ref{e42}) holds, then a system of linear (generally speaking, singular)
differential equations (\ref{e44}) has only the unique solution for each
initial state \cite{20}. Hence, since $x(t_2)=0,$ by (\ref{e44}) $x(t)\equiv
0,t_2+h_{m-1}-h_m<t\le t_2$. Proceeding in a similar way with $%
t_2+h_{m-1}-h_m$ instead $t_2$, we will obtain, that $x(t)\equiv
0,t_2+2(h_{m-1}-h_m)<t\le t_2$. Continuing the same arguments, we will
obtain in a finite number of steps, that $x(t)\equiv 0,-h\le t\le 0$, that
is, $x_0=0.$ Thus the system (\ref{e31})-(\ref{e32}) is non-pasted.

{\it Necessity}. If $\det \{A_{0m}\lambda +A_{1m}\}\equiv 0$ for all
$\lambda \in {\bf C}$ then equation
\begin{equation}
\label{e45}A_{0m}\dot \zeta (\tau )+A_{1m}\zeta (\tau )\equiv 0,-h_m<\tau
\le -h_{m-1}
\end{equation}
has a non-trivial solution $\zeta (\tau ),\zeta (-h_{m-1})=0$\cite{20}.

Let
$$
\varphi (\tau )=
\cases{0,&if $-h_{m-1\leq }\tau \leq 0,$ \cr
\zeta (\tau ),&if $-h_m\leq \tau \leq -h_{m-1}.$ \cr}
$$
It is easy to show (see the formula for solutions of (\ref{e40}) in \cite{18}%
), that in this case $x(t)\equiv 0,t\ge 0.$ Hence, $S(t)x_0\equiv 0$ for $%
t\ge h$, and the equation (\ref{e31})-(\ref{e32}) is pasted on $[0,h]$.

It is easy to show from proof of Theorem \ref{T5}, that pasted equation (\ref
{e31})-(\ref{e32}) with output (\ref{e34}) isn't observable. Hence, the next
theorem holds.

\begin{theorem}
\label{T6} For the equation (\ref{e31})-(\ref{e34}) to be observable on $%
[0,t_1]$, it is necessary and sufficient, that (\ref{e35}) and (\ref{e42})
hold.
\end{theorem}

\section{Conclusion}

In this paper a criterion of the approximate observability by means of
linear operation is obtained and formulas (\ref{e11}), (\ref{e20}), (\ref
{e28}), (\ref{e29}) for the determination of these linear operations are
established. In order to use this formulas, one should know the elements of
biorthogonal systems for sequence (\ref{e8}). In general case it is not easy
to solve this problem for a concrete sequence (\ref{e8}), and the sequence (%
\ref{e8}) must itself be determined. In many cases to determine of this
sequence it is required to solve a complicated transcendental equation \cite
{18}. If we can solve this equation and find the sequence (\ref{e8}) and
elements of the biorthogonal system, we can use formulas (\ref{e11}), (\ref
{e20}), (\ref{e28}), (\ref{e29}) in order to determine a sequence of linear
operations, which solves the approximate observability problem.

If the series
\begin{equation}
\label{e46}
U(\tau )y=-\sum_{j=1}^\infty \sum_{k=1}^{\beta _j}
(\varphi_{jk}v_{jk}^T(\tau ))y
\end{equation}
converges, then (\ref{e28}) becomes
\begin{equation}
\label{e47}x(t)=\int_0^{t_1}U(\tau )y(t-\tau )d\tau ,\forall t\ge t_1.
\end{equation}

\begin{definition}
\label{D7} 
Equation (\ref{e1})-(\ref{e3}) is called completely observable on $[0,t_1]$
in the class of linear operations, if for any each solution $x(t)$ of
equation (\ref{e1})-(\ref{e2}) there exists an operator-valued function
$$
U(t),0\le t\le t_1,U\in L_2([0,t_1],L(Y,X))
$$
such that (\ref{e47}) holds.
\end{definition}

It is possible to show, that conditions of complete observability of
equation (\ref{e1})-(\ref{e3}) on $[0,t_{1}]$ in the class of linear
operations by means of Theorem \ref{T1} are reduced to conditions of
convergence of series (\ref{e46}). These conditions was not investigated in
general case.

Theorem \ref{T4} has been proved in \cite{17} for neutral systems
(\ref{e40})-(\ref{e41}) without delays in control
$(B_{j}=0,j=1,\ldots, m)$ for $t_{1}>nh.$

Theorem \ref{T6} has been obtained in \cite{7} for retarded systems
(system (\ref{e40})-(\ref{e41}) with $A_{0j}=0,j=1,\ldots ,m$)
without delays in control.

Both last theorems were proved by methods specific for systems with
aftereffect.

a\end{document}